\documentclass[prd,showpacs,twocolumn,floats,floatfix,longbibliography]{revtex4-1}
\usepackage[utf8x]{inputenc}
\usepackage{color}
\usepackage{graphicx}
\usepackage{amssymb}
\usepackage[bookmarks, bookmarksopen, bookmarksnumbered, colorlinks]{hyperref}

\newcommand{\beq}{\begin{equation}}
\newcommand{\eeq}{\end{equation}}
\newcommand{\bea}{\begin{eqnarray}}
\newcommand{\eea}{\end{eqnarray}}
\newcommand{\bit}{\begin{itemize}}
\newcommand{\eit}{\end{itemize}}
\newcommand{\bfi}{\begin{figure}}
\newcommand{\efi}{\end{figure}}
\newcommand{\bfic}{\begin{figure*}}
\newcommand{\efic}{\end{figure*}}
\newcommand{\bce}{\begin{center}}
\newcommand{\ece}{\end{center}}
\newcommand{\bt}{\begin{table}}
\newcommand{\et}{\end{table}}
\newcommand{\btb}{\begin{tabular}}
\newcommand{\etb}{\end{tabular}}

\newcommand{\code}[1]{\texttt{#1}}
\newcommand{\CC}{C\nolinebreak\hspace{-.05em}\raisebox{.4ex}{\tiny\bf +}\nolinebreak\hspace{-.10em}\raisebox{.4ex}{\tiny\bf +}}

\begin{document}

\title{An automatically generated code for relativistic inhomogeneous cosmologies}
\author{Eloisa Bentivegna}
\email{eloisa.bentivegna@unict.it}
\affiliation{
Dipartimento di Fisica e Astronomia\\
Universit{\`a} degli Studi di Catania\\
\&\\
INFN, Sezione di Catania\\
Via S.~Sofia 64, 95123 Catania\\
Italy\\}

\begin{abstract}
The applications of numerical relativity to cosmology are on the rise,
contributing insight into such cosmological problems as structure formation,
primordial phase transitions, gravitational-wave generation, and inflation. In 
this paper, I present the infrastructure for the computation of inhomogeneous dust
cosmologies which was used recently to measure the effect of nonlinear inhomogeneity
on the cosmic expansion rate. I illustrate the code's architecture, provide evidence 
for its correctness in a number of familiar cosmological settings, and evaluate 
its parallel performance for grids of up to several billion points. The code, which
is available as free software, is 
based on the Einstein Toolkit infrastructure, and in particular leverages the 
automated-code-generation capabilities provided by its component \texttt{Kranc}.
\end{abstract}


\maketitle

\section{Introduction}
One of the most remarkable successes of general relativity is that it enables
the relatively straightforward construction of cosmological models in outstanding agreement with 
the observational data. As as result, over the past few decades, a remarkably articulate
picture of the large-scale Universe has emerged, which combines exact models,
relativistic perturbation theory and Newtonian $N$-body simulations into a 
powerful prediction tool. This is usually referred to as the {\it Concordance Model}~\cite{ellis2012relativistic}.

Over the same period, however, cosmological observations have improved in 
volume and accuracy, challenging the validity of the previous approach and 
encouraging researchers to explore fully relativistic, non-perturbative
modelling schemes. The imminent perspective of percent-accuracy datasets 
raises the question of whether the non-linear relativistic effects neglected
by the Concordance Model will soon be observable~\cite{Bull:2015stt,Debono:2016vkp},
and perhaps even serve as a testbed for modified-gravity theories~\cite{Fleury:2016tsz}.

The new schemes include relativistic corrections to $N$-body simulations
\cite{Bruni:2013mua,Thomas:2015kua,Adamek:2016zes,Adamek:2015eda,Adamek:2015hqa,Adamek:2014xba,Adamek:2014gva},
spherically-symmetric structure-formation studies using a scalar field 
coupled to Einstein's equation~\cite{Torres:2014bpa,Alcubierre:2015ipa,Rekier:2015isa,Rekier:2015kxa},
and fully 3D models of inhomogeneous relativistic dust~\cite{Giblin:2015vwq,Mertens:2015ttp,Bentivegna:2015flc,Giblin:2016mjp}.

In this work, I describe in detail the approach used in~\cite{Bentivegna:2015flc},
including the system of equations that govern the gravitational 
and matter fields, the algorithms used to solve them, and the
techniques used to implement such algorithms. I will then prove
that this scheme reproduces known cosmological solutions filled
with a pressureless fluid, such as expanding homogeneous and 
isotropic spaces, collapsing spherically-symmetric overdensities,
and expanding axisymmetric underdensities in the presence
of a cosmological constant. This implementation is available as free
software to interested researchers~\footnote{The code is distributed
under the GNU General Public License, version 3 or later (GNU GPLv3+) 
from the repository \url{https://bitbucket.org/eloisa/ctthorns.git}}.

In the next section, I will illustrate the perfect-fluid representation 
of cosmic matter fields which underpins the treatment in~\cite{Bentivegna:2015flc}.
In section~\ref{sec:code}, I will describe the software modules used to
implement such representation, while section~\ref{sec:tests} presents
the application of this infrastructure to the computation of various
well-known exact cosmologies. In section~\ref{sec:perf} I discuss its 
performance, and conclude in section~\ref{sec:conc}. All quantities are
expressed in geometric units, $G=c=1$. 
Latin letters from the beginning of the alphabet ($a,b,c,\dots$)
denote abstract indices, while those from the middle ($i,j,k,\dots$)
are used for spatial components.

\section{Perfect fluids in cosmology}
According to general relativity, the Universe's gravitational field
obeys Einstein's equation coupled to energy and momentum sources
such as massive particles or photons. At late times and in rough terms, 
this results in a homogeneous and isotropic expansion at large scales 
(say, above one Gigaparsec), with the formation of gravitational 
structures on smaller scales.

In order to model these processes exactly, one would need to integrate
Einstein's equation on a space without symmetries, with features at 
several different scales, and with no recourse to the superposition
principle as the dynamics is in principle non-linear. Even with the 
aid of supercomputers, simulating a realistic relativistic Universe
is for the moment beyond our reach.

Given the impossibility of a brute-force approach, several 
simplifications are in order. First of all,
one can construct cosmological models where individual particles
are represented in statistical terms, via a macroscopic fluid 
(typically with zero pressure) which fills the Universe. This is already
a huge shortcut, which allows us to replace many microscopical degrees 
of freedom with a few macroscopical, averaged properties such as the
density and its first few moments. The process of averaging the microscopical
degrees of freedom is of course still nontrivial~\cite{ellis2012relativistic,Andersson:2006nr}, but assuming 
we can formulate a reasonable continuum-fluid description of matter,
the complexity of the problem is greatly reduced.

We now face the problem of integrating the equations of motion for
the gravitational field and the hydrodynamical properties of an
averaged fluid, the proxy for our cosmic distribution of matter.
These equations are just as nonlinear are those we started with,
so that numerical integration is the only viable option.

Notice that further simplifications could apply to specific regimes.
As an example, since gravitational structures and the associated 
non-linearities grow with time, there is a period in our cosmic history
where the degree of inhomogeneity is so small that perturbations
can be described at linear order, 
where they can be superimposed directly and, by construction, their
behaviour is decoupled from that of the underlying spacetime.
Such a condition is, for instance, well satisfied at the recombination
era, where the Universe is filled with a mix of fluids which is homogeneous
and isotropic for a part in $10^6$, as testified by the temperature fluctuations in the
cosmic-microwave-background photons, a snapshot of the cosmic conditions at that era.
As the Universe evolves, however, the relative importance of non-linear
mode coupling in the density of the averaged fluid increases.

In general, the numerical integration of the full system of
partial differential equations describing matter and gravity is the only
viable option. This approach has been pioneered in the study of compact
stars~\cite{lrr-2008-7}. In cosmology, matter would typically be modelled as
a perfect fluid, characterized by the stress-energy tensor:
\beq
T_{ab} = \rho h u_a u_b + p g_{ab}
\eeq
where $\rho$, $h=1+\epsilon+p/\rho$, $\epsilon$, $p$, and $u_a$ are the fluid
rest-mass density, specific enthalpy, specific internal energy, pressure, and 
four-velocity, respectively. The metric tensor is represented by $g_{ab}$.

Introducing a 3+1 decomposition of this quantity, so that the four-dimensional
line element reads:
\beq
d s^2 = - \alpha^2 d t^2 + \gamma_{ij} (\beta^i dt + dx^i) (\beta^j dt + dx^j) 
\eeq
the conservation properties of the fluid can be translated into time evolution
equations. Specifically, if the fluid is to satisfy the conservation law
\beq
\nabla_a T^{ab} = 0
\eeq
along with the conservation of the baryon number
\beq
\nabla_a (\rho u^a) = 0
\eeq
then the following system of equations must hold~\cite{lrr-2008-7,PhysRevD.58.064010}:
\bea
\partial_t D &=& - \partial_i (D V^i) \label{eq:rhd1} \\
\partial_t E &=& - \partial_i (E V^i) - p\, \partial_t W - p\, \partial_i (W V^i)\\
\partial_t S_i &=& - \partial_j (S_i V^j) + \frac{S_0}{2} \partial_i g_{00} +
S^j\partial_i \beta_j \nonumber \\
&&+ \frac{S^j S^k}{2 S_{0}} \partial_i g_{jk}
- \sqrt{-g} \partial_i p \label{eq:rhd2}
\eea
where I have introduced the variables:
\bea
D &=& \rho W \label{eq:DD}\\
E &=& \rho \epsilon W \label{eq:EE}\\
S_a &=& \rho h u_a W \label{eq:SS}\\
V^i &=& \frac{u^i}{u^0} \\
W &=& \sqrt{-g} u^0
\eea 
Notice that this system does not include an equation for the pressure. In order to close
it, one has to supply an equation of state $p \equiv p(\rho,\epsilon)$. The option 
$p(\rho,\epsilon)=0$ corresponds to a pressureless fluid, often referred to as {\it dust}
in cosmological contexts.

In~\cite{Bentivegna:2015flc}, this system was complemented with 
Einstein's equation in the Baumgarte-Shapiro-Shibata-Nakamura (BSSN) formulation~\cite{Shibata:1995we,Nakamura:1987zz,Baumgarte:1998te}:
\bea
(\partial_t - \beta^l \partial_l) \psi &=& - \frac{1}{3} \alpha K + \frac{1}{3} \partial_i \beta^i \label{eq:psi}\\
(\partial_t - \beta^l \partial_l) K &=& - D_i D^i \alpha + \alpha (\bar A_{ij} \bar A^{ij} + \frac{1}{3} K^2) \\
(\partial_t - \beta^l \partial_l) \bar \gamma_{ij} &=& -2 \alpha \bar A_{ij} + 2 \bar \gamma_{i(j}\partial_{k)} \beta^i - \frac{2}{3} \bar \gamma_{ij} \partial_k \beta^k\\ 
(\partial_t - \beta^l \partial_l) \bar A_{ij} &=& \psi^2 (-D_i D_j \alpha + a R_{ij})^{TF} \nonumber \\ 
  && + \alpha (K \bar A_{ij} - 2 \bar A_{ik} \bar A^k_j) \\
  && + 2 \bar A_{k(i} \partial_{j)} \beta^k - \frac{2}{3} A_{ij} \partial_k \beta^k \nonumber \\
(\partial_t - \beta^l \partial_l) \bar \Gamma^i &=& \bar \gamma^{jk} \beta^i \partial_j \beta_k + 
    \frac{1}{3} \bar \gamma^{ij} \partial_j \partial_k \beta_k - \bar \Gamma^j \partial_j \beta^i \nonumber \\ 
  &&+ \frac{2}{3} \bar \Gamma^i \partial_j \beta^j 
      - 2 \bar A^{ij} \partial_j \alpha \label{eq:Gam} \\
  &&+ 2 \alpha (\bar \Gamma^i_{jk} \bar A^{jk} - 3 \bar A^{jk} \partial_k \ln \psi - \frac{2}{3} \bar \gamma^{ij} \partial_j K) \nonumber
\eea
where:
\bea
\gamma_{ij}   &=& \psi^{-2} \bar \gamma_{ij} \\
K_{ij}        &=& \frac{K}{3} \gamma_{ij} + \psi^{-2} \bar A_{ij} \\
\bar \Gamma^i &=& -\partial_j \bar \gamma^{ij} 
\eea
In both the study in~\cite{Bentivegna:2015flc} and the tests below,
the gauge variables $\alpha$ and $\beta^i$ are set to one and zero, respectively, at all times.

\section{Code and setup}
\label{sec:code}
Computing the geometry of a spacetime filled with a fluid requires the numerical integration of 
the equations of relativistic hydrodynamics coupled to Einstein's equation. The Einstein Toolkit~\cite{Loffler:2011ay},
an open-source infrastructure for relativistic astrophysics, provides most of the infrastructure 
necessary for this task: \code{McLachlan}~\cite{mclachlan,kranc}, 
a module for the integration of Einstein's equation according to the BSSN formalism (\ref{eq:psi}-\ref{eq:Gam}),
the AMR package \code{Carpet}~\cite{carpet,carpetweb}, and a software framework called
\code{Cactus}~\cite{cactus} that ties together all the different components, and provides 
correctness enforcement~\cite{Bentivegna:2011id} and
an interface to low-level
tasks (such as I/O and parallelization), as well as to user-provided scientific modules. The infrastructure
also contains \code{CT\_MultiLevel}, a multigrid elliptic solver that can be used for the generation
of initial data~\cite{Bentivegna:2013xna}.

In order to carry out the evolution of cosmological models with fluids, one must supply a module that integrates the corresponding
equations. Perfect fluids, for instance, are governed by the system (\ref{eq:rhd1}-\ref{eq:rhd2}). 
The Einstein Toolkit provides, via the \code{Kranc} package, a mechanism for the generation
of code representing the discretization of arbitrary first-order-in-time systems of partial differential
equations.

A new module, called \code{CT\_Dust} and presented in this paper, is constructed using this mechanism. 
Specifically, the module implements the following recipe:
\begin{itemize}
\item It computes initial conditions for the hydrodynamic variables $\rho$, $\epsilon$, $p$ and $u^a$,
and fills the stress-energy tensor with these values;
\item It converts the primitive variables $\rho$, $\epsilon$, and $u^a$ into
$D$, $E$, and $S^a$ using equations (\ref{eq:DD}-\ref{eq:SS});
\item It provides a fourth-order-in-space discretization of the system (\ref{eq:rhd1}-\ref{eq:rhd2}), and
uses it to update $D$, $E$, and $S^a$;
\item It converts these variables back into the primitive ones, and fills the stress-energy tensor with
the updated values required for the integration of Einstein's equation. 
\end{itemize}

The various operations entailed by this recipes, such as the expansion of tensorial expressions,
the discretization of derivatives, and the optional coordinate transformations,
are turned into \CC~code automatically by \code{Kranc}, a fast and reliable route 
to the implementation of complex equations and initial conditions. In order to
illustrate the advantages of automated code generation in this context, I list
some of the full expressions involved in the example of section~\ref{sec:LTB} (a collapsing,
spherically-symmetric model) in Appendix~\ref{app:ltb}. 

Two important facts are worth mentioning: first, recovering the primitive variables usually
involves a root-finding algorithm, as the relationships (\ref{eq:DD}-\ref{eq:SS}) cannot be inverted analytically.
This step is almost trivially implemented in \code{CT\_Dust} as \code{Kranc} provides an interface
to corresponding tools in \code{Mathematica}.
Second, relativistic astrophysics applications almost invariably require the 
deployment of shock-capturing discretization schemes to correctly resolve and evolve fluid
discontinuities. This procedure is necessary, for instance, in scenarios where the fluid is concentrated in
compact regions surrounded by vacuum, as would be the case for a neutron star. In the
cosmological applications described in \cite{Bentivegna:2015flc}, however, this is hardly
relevant. For this reason, at this stage \code{CT\_Dust} only employs standard centered finite-differencing.

\section{Code tests}
\label{sec:tests}
In this section, I verify that the scheme described in section~\ref{sec:code} and implemented
in \code{CT\_Dust} is able to reproduce three well-known cosmological models: the flat 
Friedman-Lema{\^i}tre-Robertson-Walker (FLRW) spacetime, a Lema{\^i}tre-Tolman-Bondi (LTB)
spherical collapse model, and an axisymmetric Szekeres model with a positive
cosmological constant. In the Ellis \& van~Elst classification of~\cite{Ellis:ee}, these
models belong to the $(s=3,q=3)$, $(s=2,q=1)$, and $(s=1,q=1)$ classes, respectively.

\subsection{A FLRW model}

The simplest application of the hydrodynamical equations illustrated above is
obtained by coupling this system to Einstein's equation and evolving maximally-symmetric
initial data, i.e.~3D spaces of constant scalar curvature. In cosmology, these
spaces and their time development are referred to as FLRW models, and depending on
the equation of state their time behavior can be integrated exactly or through
the solution of an ordinary differential equation. They are therefore ideal for a code test.

The case where the spatial curvature is null and the equation of state is
given by $p=0$ is called the Einstein-de~Sitter (EdS) solution.
I integrate the coupled system on a domain given by $-L \leq \{x,y,z\} \leq L$, with 
periodic boundary conditions at the faces and resolution $\Delta_x=\Delta_y=\Delta_z=0.1 L \equiv \Delta$. 
For initial data, I choose:
\bea
t_{\rm ini} &=& \frac{2}{3 H_{\rm ini}}\\
\gamma^{\rm ini}_{ij} &\equiv& \gamma_{ij}(t_{\rm ini})= \delta_{ij}\\
\rho_{\rm ini} &\equiv& \rho(t_{\rm ini})=\frac{3 H_{\rm ini}^2}{8 \pi}
\eea
where I have expressed all dimensional quantities in terms of the initial expansion rate
$H_{\rm ini}$. 
The volume element and density of the EdS model subsequently vary in time as:
\beq
\sqrt{\gamma/\gamma^{\rm ini}} = \left(t/t_{\rm ini} \right)^2 = \rho_{\rm ini}/\rho
\eeq
where $\gamma$ is the determinant of $\gamma_{ij}$.
Setting $H_{\rm ini}=\frac{2}{3 L}$, in particular, the initial data takes the form:
\bea
t_{\rm ini} &=& L\\
\gamma_{\rm ini} &=& 1\\
\rho_{\rm ini} &=& \frac{1}{6 \pi L^2}
\eea
Normalizing these quantities with respect to a different value of $H_{\rm ini}$
or expressing them in physical units is trivial.

I evolve this initial data using $\Delta t = 0.2 \Delta$ up to $t=1000 L$, and
plot the truncation error for $a(t)\equiv\gamma(t)^{1/6}$, $H(t)\equiv\dot{a}(t)/a(t)$, and $\rho(t)$ in Figure~\ref{fig:eds}.
I run an additional resolution $\Delta/2$, also plotted in Figure~\ref{fig:eds}.
One observes, as expected, that the numerical estimates for all plotted
quantities converge to the exact solution at fourth order in the limit
$\Delta \to 0$.
As a measure of the accumulated numerical error in this timespan, I also examine
how well the mass $\sqrt{\gamma} \rho$ of the cubic cell is conserved throughout 
the evolution. As can be seen in Figure~\ref{fig:eds}, this quantity is conserved
down to the roundoff level.
Finally, two plots of the Hamiltonian-constraint violation (unscaled and scaled
by the density) are shown, which again demonstrate fourth-order convergence
to the exact solution.

\bce
\bfic
\centering
\includegraphics[width=0.95\textwidth]{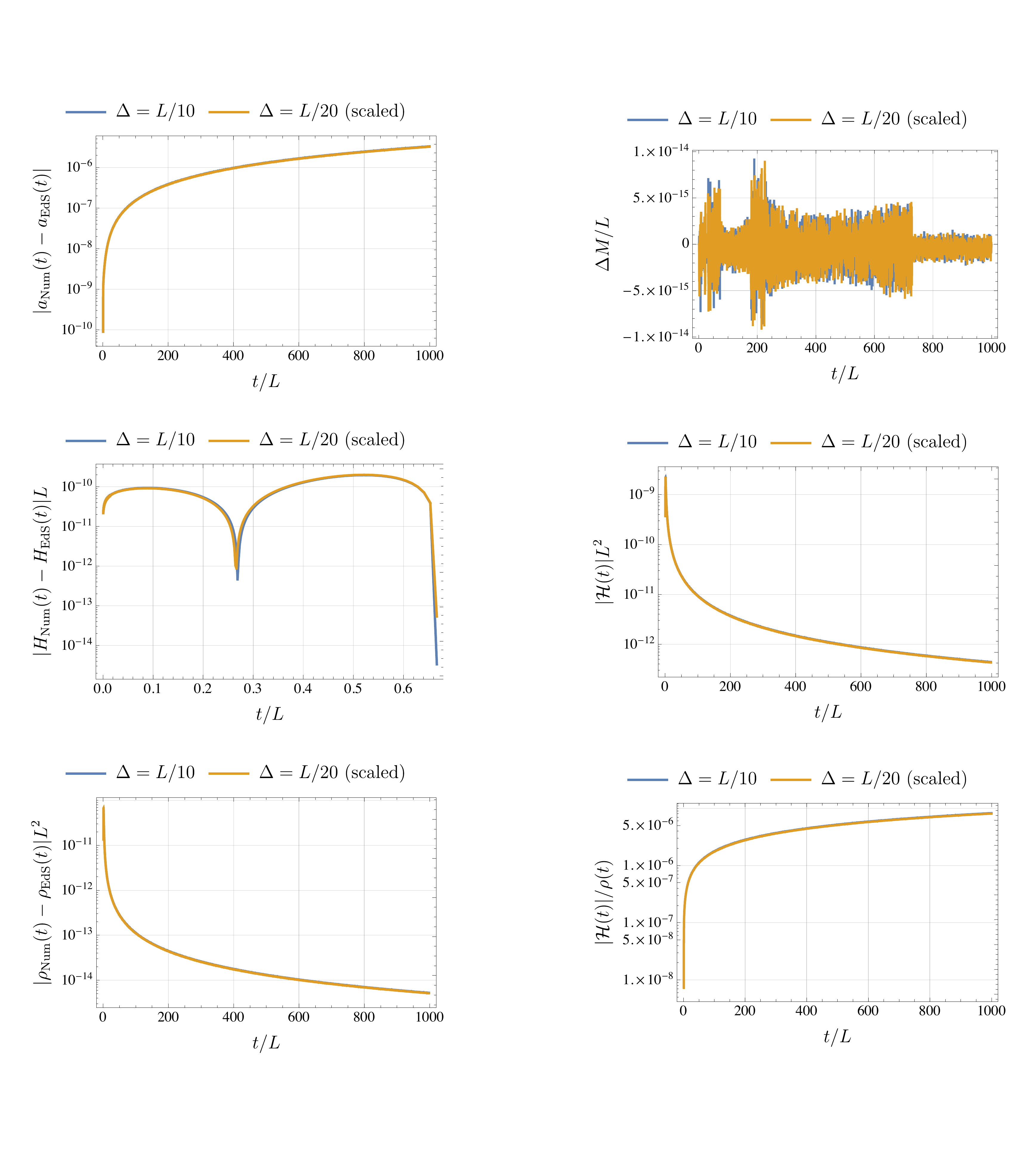}
\caption{
Left column: truncation errors in the volume element, 
expansion rate and density of the EdS solution, for two different 
resolutions $\Delta=L/10$ and $L/20$ (the latter rescaled 
by the expected factor for fourth-order convergence). Right column: 
violations in the total-mass conservation and the Hamiltonian constraint
(absolute and normalized to the density), also for two resolutions.
\label{fig:eds}}
\efic
\ece

\subsection{A LTB model}
\label{sec:LTB}
Another relevant class of cosmologies is given by the spherically-symmetric 
LTB models,
represented by the line element~\cite{Ellis:ee}:
\beq
ds^2 = - dt^2 + X^2(t,r) dr^2 + Y^2(t,r) d \Omega^2
\eeq
The corresponding metric tensor is a solution of Einstein's equation if:
\bea
X(t,r) &=& \pm \frac{|Y'(t,r)|}{1+2E(r)} \\
Y(t,r) &=& \frac{M(r)}{\epsilon(r)} \Phi_0(t,r) \\
\rho(t,r) &=& \frac{M'(r)}{4 \pi Y^2(t,r) Y'(t,r)}
\eea
where $E(r)$ and $M(r)$ are freely specifiable radial functions,
the prime denotes differentiation with respect to $r$, and:
\bea 
\epsilon(r) &=& \left\{
  \begin{array}{rr}
    2 E(r) & \qquad \textrm{for } E(r) > 0 \\
    1      & \textrm{for } E(r) = 0 \\
  - 2 E(r) & \textrm{for } E(r) < 0
  \end{array}
\right. \\
\Phi_0(t,r) &=& \left\{
  \begin{array}{ll}
    \cosh(\eta(t,r)) - 1       & \textrm{for } E(r) > 0 \\
    \eta^2(t,r)/2              & \textrm{for } E(r) = 0 \\
    1 - \cos(\eta(t,r))        & \textrm{for } E(r) < 0
  \end{array}
\right.
\eea
The function $\eta(t,r)$ is a solution of the equation
\beq
\frac{|\epsilon(r)|^{3/2}(t-t_{\rm B}(r))}{M(r)} = \xi(\eta(t,r))
\eeq
where:
\beq
\xi(\eta) = \left\{
  \begin{array}{ll}
    \sinh \eta  - \eta  & \qquad \textrm{for } E(r) > 0 \\
    \eta^3/6            & \qquad \textrm{for } E(r) = 0 \\
    \eta - \sin \eta    & \qquad \textrm{for } E(r) < 0
  \end{array}
\right.
\eeq

This general form can describe models with constant-time spaces of any curvature,
depending on the value of $E(r)$. Notice that the metric tensor is degenerate
on the curves $(r,t_B(r))$; like $E(r)$ and $M(r)$, the function $t_B(r)$ can 
also be chosen arbitrarily, allowing one to construct models with space-dependent
Big Bangs or Big Crunches.

Here, I choose a so-called ``parabolic'' (i.e., $E(r)=0$) model where:
\beq
t_B(r)=\tilde t-\frac{1}{1+(r/\tilde r)^2}
\eeq
and the mass function profile is set to the EdS value:
\beq
M(r)=\frac{2}{9} r^3
\eeq
The corresponding expressions for the metric tensor and the matter density
in both polar and cartesian coordinates can be found in Appendix~\ref{app:ltb}.

Much like in the previous section, there is an overall length scale $L$ 
which can be set freely.
I use the above expressions to set the initial conditions at $t=L$ in the box
$-L \leq \{x,y,z\} \leq L$, with spatial resolutions $\Delta = L/40$ and 
$\Delta/2$ and boundary conditions set to the analytic solution. 
I also use $\tilde t = 5 L$ and $\tilde r = \sqrt{1/10}L$.

With the chosen parameters, the model will incur a
curvature singularity at the origin at $t=4L$; correspondingly, I observe 
that the density grows unbounded and the volume element
shrinks to zero at this point. The evolution is otherwise well behaved and
close to the analytical model, to which it converges to fourth order.
I illustrate the spatial profiles of the density, the metric component
$g_{xx}$, and the extrinsic curvature component $K_{xx}$ along the $z$
axis in the first column of Figure~\ref{fig:ltb}. In the second column,
the convergence of some of the fields at the representative point $P=(1/2,1/2,1/2)$ as a function of
time is shown.

\bce
\bfic
\centering
\includegraphics[width=0.95\textwidth]{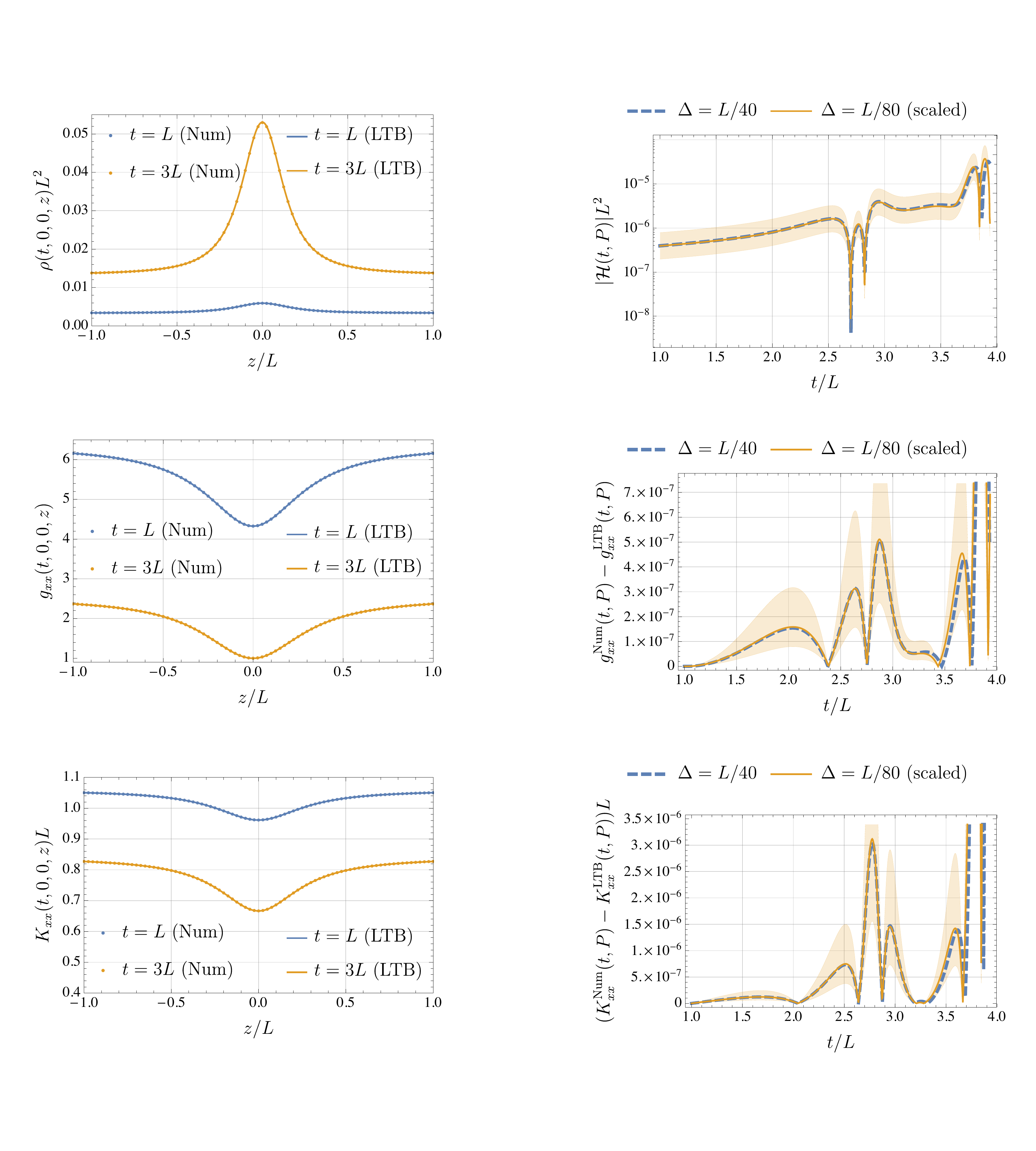}
\caption{
Spatial profiles and temporal evolution for a collapsing LTB model.
Left column: spatial profiles for $\rho$, $g_{xx}$, and $K_{xx}$ on 
the $z$ axis initially ($t=L$) and at $t=3 L$. Right column: convergence
plot for the Hamiltonian constraint violation $\cal{H}$, $g_{xx}$, and $K_{xx}$
as a function of time at a point $P=(1/2,1/2,1/2)$. The blue dashed line represents the truncation
error for the coarser resolution $\Delta=L/40$, while the thicker yellow line is
the truncation error for the finer resolution $\Delta=L/80$, rescaled for fourth-order
convergence. The thinner yellow lines represent the scaling expected
for third- and fifth-order convergence.
\label{fig:ltb}}
\efic
\ece

\subsection{A Szekeres model}

One can further release the symmetry assumptions by considering models
invariant under the action of a one-dimensional isotropic group~\cite{Ellis:ee}, such as the 
Szekeres class of models analyzed by Meures \& Bruni in~\cite{Meures:2011ke}.

In this spacetime, the density can be made axisymmetric (say, around the $z$-axis)
with an arbitrary profile along $z$. The line element can be written as:
\beq
ds^2 = -dt^2 + S(t)^2 \left[ dx^2 + dy^2 + Z(t,z)^2 dz^2\right]
\eeq
where
\bea
&&S(t) = \left ( \frac{1-\Omega_\Lambda}{\Omega_\Lambda} \right)^{1/3} \sinh^{2/3}
\left [\frac{3}{2} H_{\rm ini} \sqrt{\Omega_\Lambda} (t+t_\star) \right ] \\
&&Z(t,z) = 1 + (1-\sin kz )[f_+(t+t_\star)+B(x^2+y^2)] 
\eea
Here, $\Lambda$ and $\Omega_\Lambda$ are the cosmological constant and its associated density
parameter, $k$ is an arbitrary wave number, $B$ is given by:
\beq
B=\frac{3}{4} H_{\rm ini}^2 \left [ \Omega_\Lambda (1-\Omega_\Lambda)^2 \right ] ^{1/3}
\eeq
and $f_+(t)$ is a solution of:
\beq
f'' + \frac{4}{3} \coth \left (\sqrt{\frac{3 \Lambda}{4}} t \right ) f' - \frac{2}{3} \frac{1}{\sinh^2 \left (\sqrt{\frac{3 \Lambda}{4}} t \right )} f = 0
\eeq

\bce
\bfic[!t]
\centering
\includegraphics[width=0.95\textwidth]{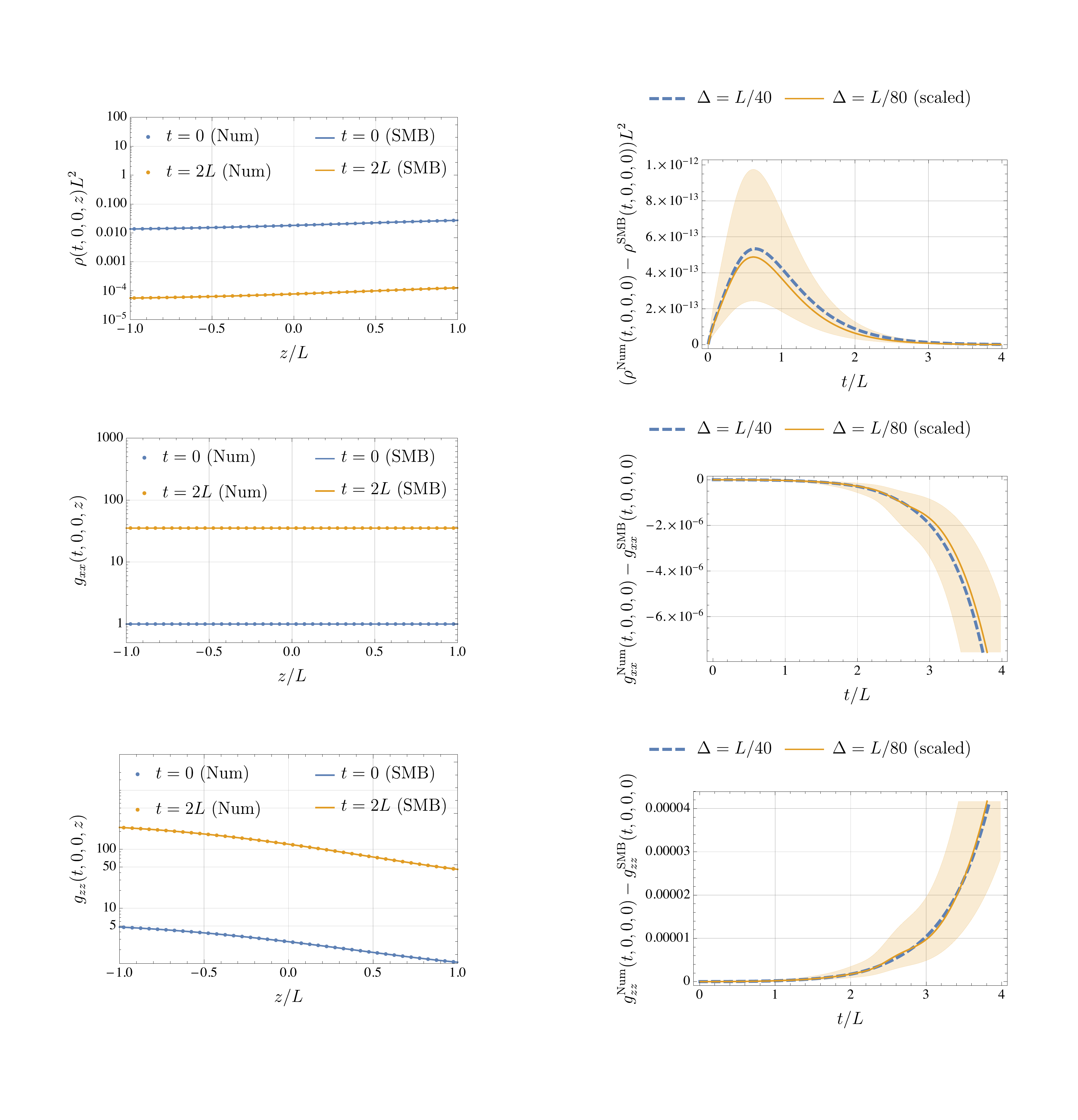}
\caption{
Left column: density and metric components $g_{xx}$ and $g_{zz}$ along
the $z$-axis at $t=0$ and $t=2L$, for a grid spacing $\Delta=L/40$. 
Right column: truncation errors for the above quantities and two 
resolutions $\Delta=L/40$ and $L/80$.
The blue dashed line represents the truncation
error for the coarser resolution, while the thicker yellow line is
the truncation error for the finer resolution, rescaled for fourth-order
convergence. The thinner yellow lines represent the scaling expected
for third- and fifth-order convergence.
\label{fig:mb}}
\efic
\ece
I set initial conditions for this model in a domain $-L \leq {x,y,z} \leq L$
on the hypersurface at $t=0$, with $\Lambda=2.25 L^2$ (so that $\Omega_\Lambda=0.75$), $H_{\rm ini}=L^{-1}$, $k=\pi$,
$t_\star=2 \sinh^{-1}(\sqrt{\Omega_\Lambda/(1-\Omega_\Lambda)})/(3 H_{\rm ini} \sqrt{\Omega_\Lambda})$; 
this is then evolved in time, applying boundary conditions 
from the exact solution at all times.
Figure~\ref{fig:mb} illustrates the spatial profile and the convergence results for
this model. Again, the results exhibit fourth-order convergence to the exact solution.

\section{Performance and scaling}
\label{sec:perf}
Even the highest-resolution simulations described in section~\ref{sec:tests}
can be run on a laptop with a moderate amount of RAM. Table~\ref{tab:160}
shows a typical throughput for runs in this range.
\bce
\bt
\caption{Run performance on a 2.9GHz Intel Core i5 laptop 
with 16GB of RAM.\label{tab:160}}
\btb{rrr}
\hline
Number of $\qquad$ & Memory $\qquad$  & Run speed   \\ 
points $N$ $\qquad$ &  (GB)  $\qquad$  & ($L$/hour) \\ 
\hline
$47^3$   $\qquad$ & 0.19 $\qquad$   & 28   \\
$87^3$   $\qquad$ & 1.2  $\qquad$   & 2.1  \\
$167^3$  $\qquad$ & 8.7  $\qquad$   & 0.15 \\
\hline
\etb
\et
\ece
As shown in Figure~\ref{fig:scalS},
the memory required by each simulation and the execution speed roughly scale
as $N$ and $N^{-4/3}$ respectively, as expected.
\bce
\bfi[!b]
\centering
\includegraphics[width=0.4\textwidth]{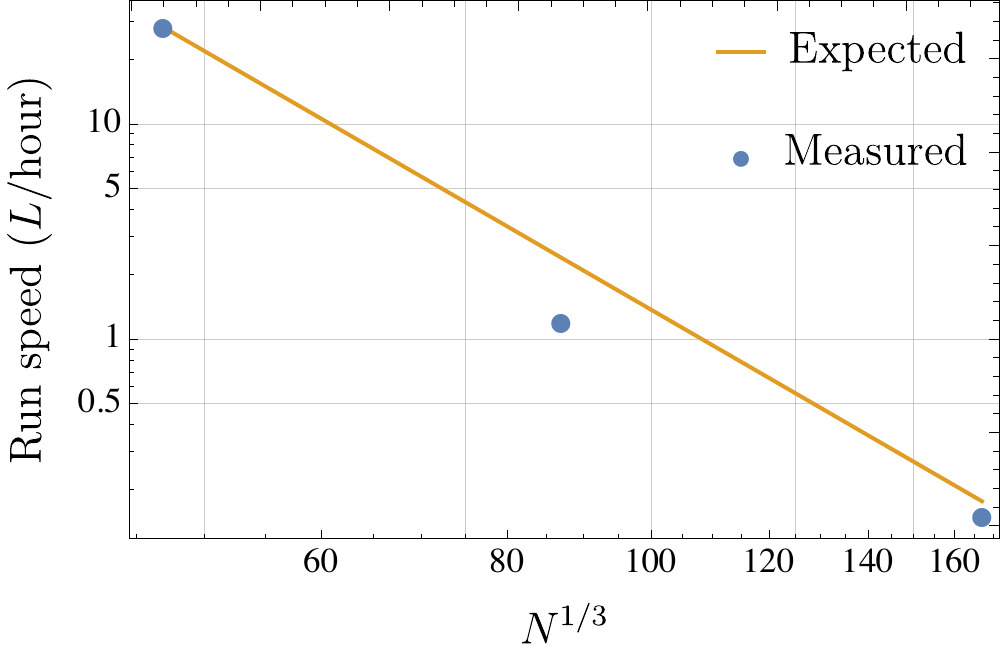}
\caption{
Run speed for grid sizes with $N=46^3, 86^3, 166^3$, normalized to
the speed at $N=46^3$. The reference curve is given by $N^{-4/3}$.
\label{fig:scalS}}
\efi
\ece
Thanks to its parallel capabilities, the code can also leverage a much 
higher number of processors, allowing for larger grid sizes. The weak- and
strong-scaling properties for grids of up to several billion points are shown in 
tables~\ref{tab:ws}-\ref{tab:ss} and Figures~\ref{fig:scalLW}-\ref{fig:scalLS}, 
for runs carried out on the Marconi system at CINECA.
The data shows that, keeping the number of points per processing core constant,
the problem size can be scaled up to $1447^3$ points, with a degradation 
in run speed of at most $8.1\%$. Furthermore, the solution of a fixed-size problem 
can be accelerated by a factor of $28$, with a parallel efficiency which is never below $76\%$. 
It is worth noting that the range of these tests was solely limited by the maximum 
job size allowed on Marconi.

\bce
\bt
\caption{Weak scaling test on CINECA's Marconi supercomputer.
Each node contains 36 processing cores.\label{tab:ws}}
\btb{rrrr}
\hline
Number of $\qquad$  & Memory $\qquad$  & Number of $\qquad$ & Run speed \\ 
points $N$ $\qquad$ &  (TB)  $\qquad$  & nodes $n$   $\qquad$ & ($L$/ hour)   \\ 
\hline
$647^3$   $\qquad$ & 0.54 $\qquad$   &  6 $\qquad$ & 0.037  \\
$807^3$   $\qquad$ & 1.1  $\qquad$   & 15 $\qquad$ & 0.036  \\
$967^3$   $\qquad$ & 1.8  $\qquad$   & 30 $\qquad$ & 0.037  \\
$1127^3$  $\qquad$ & 2.9  $\qquad$   & 56 $\qquad$ & 0.035  \\
$1287^3$  $\qquad$ & 4.3  $\qquad$   & 94 $\qquad$ & 0.034  \\
$1447^3$  $\qquad$ & 8.7  $\qquad$   &151 $\qquad$ & 0.034  \\
\hline
\etb
\et
\ece

\bce
\bfi
\centering
\includegraphics[width=0.4\textwidth]{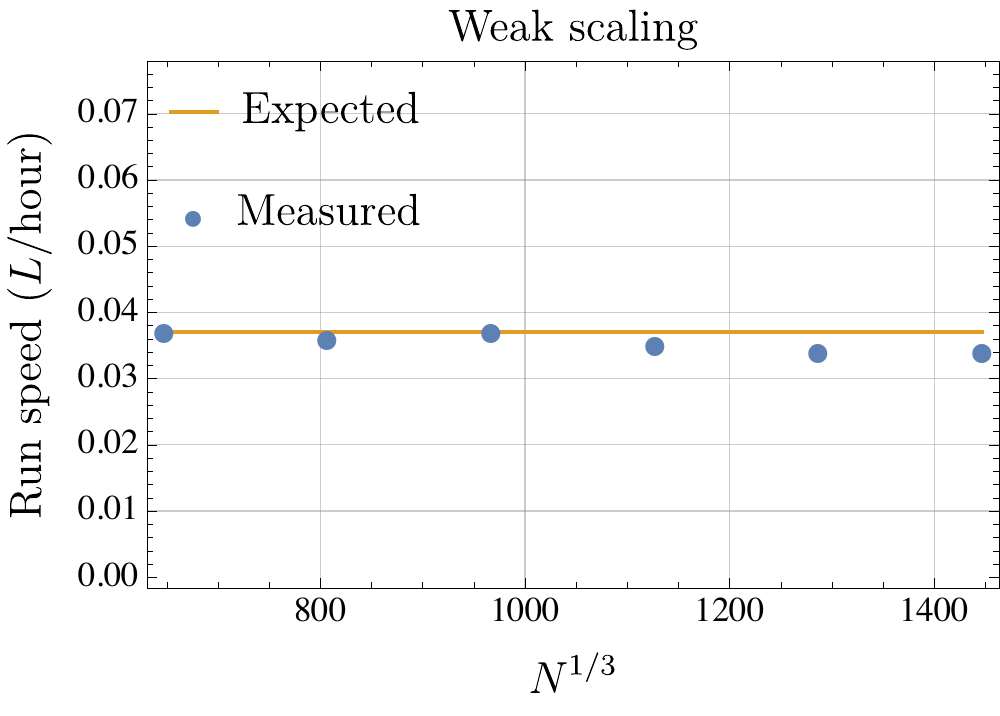}
\caption{
Weak scaling test for the simulations in Table~\ref{tab:ws}.
\label{fig:scalLW}}
\efi
\ece

\bce
\bt
\caption{Strong scaling test on CINECA's Marconi supercomputer.
Each node contains 36 processing cores.\label{tab:ss}}
\btb{rrr}
\hline
Number of $\qquad$  & Number of $\qquad$ & Run speed $\qquad$   \\ 
points $N$ $\qquad$ & nodes $n$   $\qquad$ & ($L$/hour)   $\qquad$   \\ 
\hline
$647^3$   $\qquad$ &   6 $\qquad$ & 0.037 $\qquad$ \\
$647^3$   $\qquad$ &   9 $\qquad$ & 0.054 $\qquad$ \\
$647^3$   $\qquad$ &  12 $\qquad$ & 0.068 $\qquad$ \\
$647^3$   $\qquad$ &  15 $\qquad$ & 0.085 $\qquad$ \\
$647^3$   $\qquad$ &  30 $\qquad$ & 0.17  $\qquad$ \\
$647^3$   $\qquad$ & 100 $\qquad$ & 0.50  $\qquad$ \\
$647^3$   $\qquad$ & 166 $\qquad$ & 0.78  $\qquad$ \\
\hline
\etb
\et
\ece

\bce
\bfi
\centering
\includegraphics[width=0.4\textwidth]{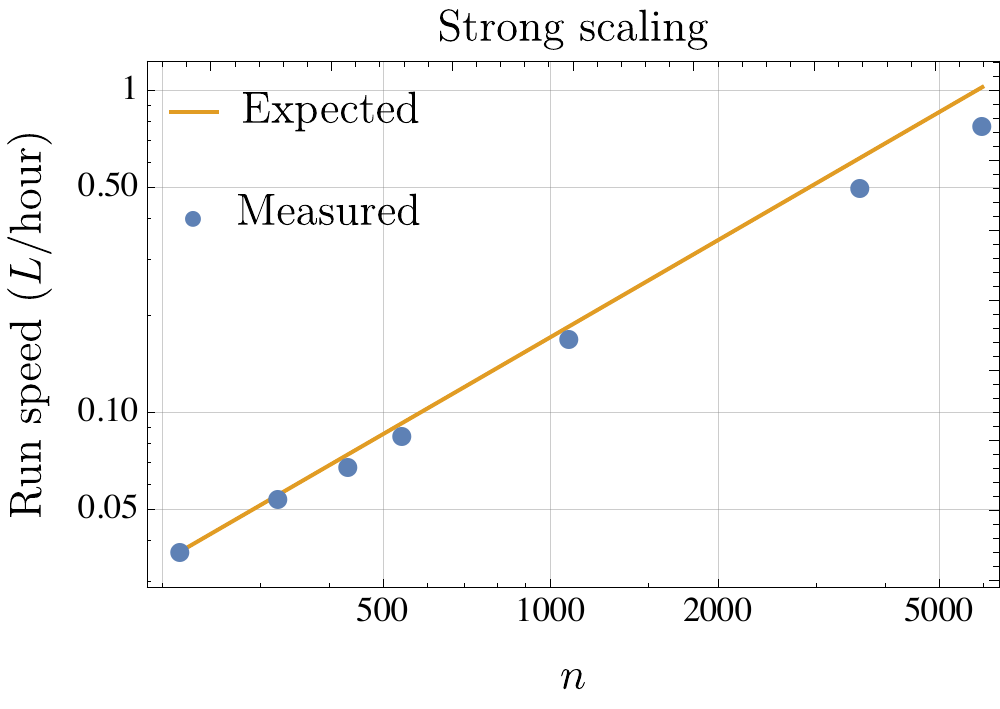}
\caption{
Strong scaling test for the simulations in Table~\ref{tab:ss}.
The slope of the reference curve is one.
\label{fig:scalLS}}
\efi
\ece

\section{Conclusions}
\label{sec:conc}
I have described a free-software infrastructure for the computation of 
inhomogeneous cosmologies through the integration of Einstein's equation
coupled to the equations of relativistic hydrodynamics. The first 
physical results obtained by this infrastructure have appeared 
in~\cite{Bentivegna:2015flc}.
This companion paper details the algorithms that form the code, provides 
verification of its behavior in known scenarios, and outlines its computational 
performance.  

As illustrated in~\cite{Bentivegna:2015flc,Giblin:2015vwq,Giblin:2016mjp,Mertens:2015ttp},
the ability to model the relativistic effects at play in the large-scale Universe
opens the road to a better comprehension of its evolution, content, and appearance.
The physics of this system is not only one of the core cultural successes of the 
theory of general relativity, but also a basic ingredient in the study of many other 
phenomena of astrophysical and cosmological nature. The infrastructure 
described in this work will surely enable many of the future studies in this 
direction.

\begin{acknowledgments}
The author is grateful to the Einstein Toolkit contributors for
their work on some of the components used in this study. 
Financial support from a Rita Levi Montalcini grant,  
``\emph{Digitizing the universe: precision modelling for precision cosmology}'', 
funded by the Italian Ministry of Education, University and Research (MIUR), is
acknowledged. 
Some of the computations were carried out on the Marconi cluster at CINECA.
\end{acknowledgments}

\appendix
\section{The LTB spacetime in Cartesian coordinates}
\label{app:ltb}
In polar coordinates, the non-zero components of the metric tensor of the
LTB model described in section~\ref{sec:LTB} take the following form:
\begin{widetext}
\bea
g_{rr}(t,r) &=& \frac{\left(3 r^4 (t-\tilde{t})+r^2 \tilde{r}^2 (6 t-6 \tilde{t}-1)+3 \tilde{r}^4 (t-\tilde{t}+1)\right)^2}{9 
   \left(r^2+\tilde{r}^2\right)^4 \left(-\frac{\tilde{r}^2}{r^2+\tilde{r}^2}-t+\tilde{t}\right)^{2/3}} \\ 
g_{\theta\theta}(t,r) &=& r^2 \left(-\frac{\tilde{r}^2}{r^2+\tilde{r}^2}-t+\tilde{t}\right)^{4/3} \\
g_{\phi\phi}(t,r) &=& \left(x^2+y^2\right) \left(-\frac{\tilde{r}^2}{r^2+\tilde{r}^2}-t+\tilde{t}\right)^{4/3}
\eea
\end{widetext}
whilst the density is given by:
\begin{widetext}
\beq
\rho(t,r) = \frac{\left(r^2+\tilde{r}^2\right)^3}{2 \pi  \left(r^2 (t-\tilde{t})+\tilde{r}^2 (t-\tilde{t}+1)\right) \left(3 r^4 (t-\tilde{t})+r^2
   \tilde{r}^2 (6 t-6 \tilde{t}-1)+3 \tilde{r}^4 (t-\tilde{t}+1)\right)}
\eeq
\end{widetext}
The same quantities are given, in Cartesian coordinates, by:
\begin{widetext}
\bea
g_{xx}(t,x,y,z) &=& r^{-4} \left ( r^2 \left(y^2+z^2\right) \left(-\frac{\tilde{r}^2}{r^2+\tilde{r}^2}-t+\tilde{t}\right)^{4/3} + \right . \nonumber \\
       && \frac{x^4 \left(3 r^4
   (t-\tilde{t})+r^2 \tilde{r}^2 (6 t-6 \tilde{t}-1)+3 \tilde{r}^4 (t-\tilde{t}+1)\right)^2}{9 \left(r^2+\tilde{r}^2\right)^4
   \left(-\frac{\tilde{r}^2}{r^2+\tilde{r}^2}-t+\tilde{t}\right)^{2/3}} + \nonumber \\
       && \left . \frac{x^2 \left(y^2+z^2\right) \left(3 r^4 (t-\tilde{t})+r^2
   \tilde{r}^2 (6 t-6 \tilde{t}-1)+3 \tilde{r}^4 (t-\tilde{t}+1)\right)^2}{9 \left(r^2+\tilde{r}^2\right)^4
   \left(-\frac{\tilde{r}^2}{r^2+\tilde{r}^2}-t+\tilde{t}\right)^{2/3}} \right ) \\
g_{xy}(t,x,y,z)  &=& -\frac{8 \tilde{r}^2 x y \left(3 r^4 (t-\tilde{t})+r^2 \tilde{r}^2 (6 t-6 \tilde{t}+1)+3 \tilde{r}^4 (t-\tilde{t}+1)\right)}{9
   \left(r^2+\tilde{r}^2\right)^4 \left(-\frac{\tilde{r}^2}{r^2+\tilde{r}^2}-t+\tilde{t}\right)^{2/3}} \\ 
\rho(t,x,y,z)  &=& \frac{\left(r^2+\tilde{r}^2\right)^3}{2 \pi  \left(r^2 (t-\tilde{t})+\tilde{r}^2 (t-\tilde{t}+1)\right) \left(3 r^4 (t-\tilde{t})+r^2
   \tilde{r}^2 (6 t-6 \tilde{t}-1)+3 \tilde{r}^4 (t-\tilde{t}+1)\right)}
\eea
\end{widetext}
where $r=\sqrt{x^2+y^2+z^2}$ and, due the spherical symmetry, the other components of 
the metric can be obtained by permuting the spatial coordinates. The computation
of these quantities and their derivatives is greatly simplified by the ability to 
generate code automatically.

\bibliography{refs}

\end{document}